\documentclass[traditabstract]{aa}
\usepackage[latin1]{inputenc}
\usepackage{graphicx}
\usepackage[english]{babel}
\usepackage{amsmath}
\usepackage{amssymb}
\usepackage{amsfonts}

\usepackage{txfonts}

\def\be{\begin{equation}}
\def\ee{\end{equation}}
\def\bea{\begin{eqnarray}}
\def\eea{\end{eqnarray}}
\def\beq{\begin{eqnarray}}
\def\eeq{\end{eqnarray}}

\begin{document}
\title{\textbf{Short Gamma Ray Bursts as possible electromagnetic counterpart of coalescing binary systems}}
\author{Salvatore Capozziello$^{1,2}$, Mariafelicia  De Laurentis$^{1,2}$, Ivan De Martino$^{1}$, Michelangelo Formisano$^{1}$}
\institute{\it $^1$Dipartimento di Scienze Fisiche, Università di
Napoli {}``Federico II'', $^2$INFN Sez. di Napoli, Compl. Univ. di
Monte S. Angelo, Edificio G, Via Cinthia, I-80126, Napoli, Italy}
\titlerunning{ Short GRBs and coalescing binaries}
\authorrunning{S. Capozziello et al.}
\abstract{ Coalescing binary systems, consisting of two collapsed
objects, are among the most promising sources of high frequency
gravitational waves signals detectable, in principle, by
ground-based interferometers. Binary systems of Neutron Star or
Black Hole/Neutron Star mergers should also  give rise to short
Gamma Ray Bursts, a subclass of Gamma Ray Bursts. Short-hard-Gamma
Ray Bursts might thus provide a powerful way to infer the merger
rate of two-collapsed object binaries. Under the hypothesis that
most short Gamma Ray Bursts originate from binaries of Neutron
Star or Black Hole/Neutron Star mergers, we outline here the
possibility to associate short Gamma Ray Bursts as electromagnetic
counterpart of coalescing binary systems.}


 \maketitle

\section{Introduction}
\label{uno}

Coalescing binary systems containing two collapsed objects, i.e. a
Neutron Star (NS), a stellar-mass Black Hole and a Neutron Star
(BH-NS) or two stellar-mass Black Holes (BH-BH) or White
Dwarf-White Dwarf (WD-WD), are supposed to emit a powerful rate of
gravitational waves (GWs). They are considered among the most
promising GWs sources for ground-based interferometers, of the
current and future generation, such as LIGO  (Laser Interferometer
Gravitational-Wave Observatory) (\cite{LIGO}) and French/Italian
VIRGO (\cite{VIRGO}) and their advanced versions. BH-NS and,
especially, BH-BH mergers emit more powerful GWs than NS mergers,
for which the sensitivity of LIGO and VIRGO detectors is highest:
therefore they can be detected up to larger distances. The horizon
of first generation LIGO and VIRGO for NSs, BH-NS and BH-BH
mergers is $\sim 20, 43$ and $100$ Mpc, respectively, while
ADVANCED LIGO/VIRGO class interferometers should detect them up to
distance of $\sim300, 650$ and $1600$ Mpc. The rate estimated of
detectable merging events has been based on the observed galactic
population of NS binaries containing a radio pulsar (\cite{phi};
\cite{Burgay}; \cite{Kalogera}; \cite{Narayan}). The best estimate
of the NS merger rate in the Galaxy is presently $\sim 80_{-70}^{
+ 210}$ Myr$^{-1}$, converting to $\sim 800^{+2000}_{-600}$
Gpc$^{-3}$ yr$^{-1}$ for a galaxy number density of
$10^{-2}$Mpc$^{-3} $(\cite{Kalogera}). The study of binary systems
population  gives results consistent with the above rate
(\cite{perna}; \cite{Belczynski}; \cite{Belczynski1}). GWs signals
from NS mergers are expected at a rate of one in $\sim10 \div150$
years with VIRGO and LIGO and one every $1\div 15$ days with
ADVANCED LIGO/VIRGO class interferometers. The BH-NS and BH-BH
merger rates in the Galaxy are highly uncertain (\cite{perna}),
they are estimated $\sim 1\%$ and $\sim0.1\%$ of the NS merger
rate, respectively, implying that BH-NS and BH-BH mergers
contribute marginally to the GWs event rate, despite the larger
distance up to which they can be detected. Recent observations
support the hypothesis that a large fraction of short Gamma Ray
Bursts (GRBs) are associated with the inspiral and merger of
compact binaries. GRBs are powerful flashes of $\gamma$-rays which
occur approximately once per day and are isotropically distributed
over the sky~(\cite{Meszaros:2006rc}). The variability of the
bursts on time scales as short as a millisecond indicates that the
sources are very compact, while the identification of host
galaxies and the measurement of redshifts for more than 100 bursts
have shown that GRBs are of extra-galactic origin. GRBs are
grouped into two broad classes by their characteristic duration
and spectral hardness~(\cite{ck93}; \cite{Gehrels2006}).  The
progenitors of most short GRBs (duration $\lesssim$ 2 s, with hard
spectra) are widely thought to be mergers of NS binaries or NS-BH
binaries; see (\cite{nakar-2007}) and references therein.  A small
fraction (up to $\simeq$15\%) of short-duration GRBs are also
thought to be due to giant flares from a local distribution of
soft-gamma repeaters (SGRs) (\cite{duncan92}; \cite{NaGaFo:06};
\cite{Chapman:2007xs}). Long GRBs (duration $\gtrsim$ 2 s, with
soft spectra), on the other hand, are associated with
core-collapse Supernovae (\cite{galama98}; \cite{hjorth03};
\cite{Ma_etal:04}; \cite{Campana:2006qe}). Both the merger and
Supernovae scenarios result in the formation of a stellar-mass
Black Hole with accretion disk (\cite{fryer99};
\cite{Cannizzo:2009qv}), and the emission of gravitational
radiation is expected in this process.

Since, as we said above, GWs measurements of well-localized
inspiraling binaries can measure absolute source distances with
high accuracy, simultaneous observation of GWs (emitted by binary
systems) and short GRBs would allow us to directly and
independently determine both the luminosity distance and the
redshift of the binary systems (\cite{Formisano}). It has long
been argued that NS-NS and NS-BH mergers are likely to be
accompanied by a Gamma Ray Burst (\cite{eichler89}).  Recent
evidence supports the hypothesis that many short-hard GRBs are
indeed associated with such mergers (\cite{fox05};
\cite{berger07}).  This suggests the exciting possibility that it
may be possible to simultaneously measure, for coalescing binary
system, the GRBs and GWs.  The combined electromagnetic and
gravitational view of these objects is likely to teach us
substantially more than what we learn from either data channel
alone. In this paper, we discuss the possibility to search for GWs
associated with 11 short GRBs that have been recently detected by
the SWIFT satellite (\cite{swift}).  In other words, we search for
GWs from short-duration GRBs. Since the nature of the radiation
depends on the somewhat-unknown progenitor model, and we analyze
short  GRBs, the search methods presented here require specific
knowledge of the gravitational waveforms which is specifically
targeting binary inspiral GW signals associated with short GRBs.
 We look for  burst signals with duration $\lesssim 2 $ s
and frequencies in the LIGO/VIRGO band, approximately
$100\div1000$ Hz. \vspace{3.mm} The paper is organized as follows.
In Sec. \ref{due}, we discuss  some GRB energy relations while, in
Sec. \ref{tre}, the energy loss by binary coalescing system is
considered.  In Sec. \ref{quattro}, the chirp mass  is calculated
by considering a sample of short GRBs in the SWIFT catalogue.
Sect. \ref{cinque} is devoted to discussion and conclusions.

\section{Gamma Ray Bursts Energy Relations}
\label{due}

GRBs are the most powerful explosions observed in the Universe
(\cite{Meszaros:2006rc}). They are believed to be detectable up to
a very high redshift (GRB090423 has been detected at redshift
$z\sim8$).

One of characteristic parameters of GRBs is T$_{90}$, which is the
time  interval within which 90$\%$ of the flux is detected. GRBs
can be roughly  separated into two classes (\cite{Weeeks}): long
GRBs (with T$_{90}$$>2$s), generally associated to Supernovae or
Hypernovae and short GRBs (with T$_{90}$$<2$s), generally
associated to mergers of compact objects.

Another  characteristic parameters is $E_p$, which is the peak
energy of the spectrum. Its measurement is available for only
cutoff power law  spectrum as the best fit value (\cite{Racusin}).
If we cannot measure $E_p$ then we can use the relation between
$E_p$ and the power law spectral index $\alpha_{PL}$
(\cite{Racusin}):
\begin{equation}\label{correl}
    \log E_p = 2.76 -3.61 \log (-\alpha_{PL}).
\end{equation}
Another interesting parameters is the collimation-corrected
energy, $E_\gamma$, that is linked to  peak energy, $E_p$,  by the
so-called  Ghirlanda relation (\cite{GGL}):
\begin{equation}\label{logEgamma}
\log E_\gamma   = a + b\log \left[ {\frac{{E_p }}{{300\rm{keV}}}} \right]
\end{equation}
where $a$ and $b$ are calibration constants (\cite{Liang}), shown in Table \ref{tab:parametri}.
\begin{table}[htbp]
\centering
\begin{tabular}{l|l|l}
\hline
\hline
\multicolumn{1}{c}{\textbf{Relation}} &
\multicolumn{1}{c}{\textbf{a}} &
\multicolumn{1}{c}{\textbf{b}} \\
\hline
$E_\gamma - E_p$ & $52.26\pm0.09$ & $1.69\pm0.11$\\
$E_{iso}-E_p-t_b$ & $52.83\pm0.10$ & $2.28\pm0.30$ \\
 &  & $-1.07\pm0.21$\\
\hline
\end{tabular}
\caption{\footnotesize{Parameter values obtained by Liang et al. 2008.}}\label{tab:parametri}
\end{table}

We  want to investigate the link between short GRBs and coalescing
binary systems, and we want show that, if the
collimation-corrected energy is correlated to the energy of GWs
(and if they are comparable), then the chirp masses associated to
short GRBs are compatible with those associated to the coalescing
binary systems.

  \section{Energy loss from binary coalescing system}\label{tre}

The energy $dE$ carried by a GW along its direction of propagation
per area $dA$ per time $dt$ is given by:
\begin{equation}
\frac{{dE}}
{{dAdt}} \equiv F = \frac{{c^3 }}
{{16\pi G}}\left[ {\left( {\frac{{\partial h_ +  }}
{{\partial t}}} \right)^2  + \left( {\frac{{\partial h_ \times  }}
{{\partial t}}} \right)^2 } \right],
\end{equation}
where $h_+$ and $h_{\times}$ are the standard polarizations of
GWs. The energy output $dE/dt$ from a localized source in all
directions is given by the integral (\cite{maggiore}):
\begin{equation}
\frac{{dE}}
{{dt}} = \int {F\left( {\theta ,\varphi } \right)} r^2 d\Omega.
\label{eq:energy}
\end{equation}
Replacing
\begin{equation}
\left( {\frac{{\partial h_ +  }}
{{\partial t}}} \right)^2  + \left( {\frac{{\partial h_ \times  }}
{{\partial t}}} \right)^2  = 4\pi ^2 f^2 h^2 \left( {\theta ,\varphi } \right)
\end{equation}
and introducing
\begin{equation}
h^2  = \frac{1}
{{4\pi }}\int {h^2 } \left( {\theta ,\varphi } \right)d\Omega,
\end{equation}
we write eq.\eqref{eq:energy} in the form
\begin{equation}
\frac{{dE}}
{{dt}} = \frac{{c^3 }}
{G}\left( {\pi f} \right)^2 h^2 r^2.
\label{eq:energy2}
\end{equation}
Specifically, for a binary system in a circular orbit, with the
help of the eq.\eqref{eq:energy2} and the definition of the
characteristic amplitude of  GW (\cite{maggiore}), we have
\begin{equation}\label{eq:amplitude}
h = \left( {\frac{{32}} {5}} \right)^{1/2} \frac{{G^{5/3} }} {{c^4
}}\frac{{\mathcal{M}^{5/3}_c }} {r}\left( {\pi f} \right)^{2/3}\,.
\end{equation}
The energy loss of the system is then
\begin{equation}
\frac{{dE}}
{{dt}} =   \left( {\frac{{32}}
{5}} \right)^{1/2} \frac{{G^{7/3} }}
{{c^5 }}\left( {\mathcal{M}_c\pi f} \right)^{10/3},
\label{eq:loss}
\end{equation}
where $\mathcal{M}_c$ is the chirp mass, $r$ is the distance from
the source and $f$ is the frequency of the GW. Introducing the
correction for the redshift, we can write  eq.\eqref{eq:loss} as:
\begin{equation}
\frac{{dE}}
{{dt}} =   \left( {\frac{{32}}
{5}} \right)^{1/2} \frac{{G^{7/3} }}
{{c^5 }}\left( {\mathcal{M}_c\pi f} \right)^{10/3} \left( {\frac{1}
{{1 + z}}} \right)^{20/3}.\label{eq:loss2}
\end{equation}
This expression is the same  that one can obtain directly from the
quadrupole formula (\cite{scott}). Since energy and angular
momentum are continuously away by the gravitational radiation, the
two masses in orbit spiral towards each other, thus increasing the
orbital frequency $\omega$. The GW frequency $f=\omega/\pi$ and
the GW amplitude $h$ are also increasing functions of time. The
rate of the frequency change is then
\begin{equation}
\dot f = \left( {\frac{{96}}
{5}} \right)\frac{{G^{5/3} }}
{{c^5 }}\pi ^{8/3} \mathcal{M}^{5/3}_c f^{11/3}.
\end{equation}
We have now all the ingredients to test our hypothesis.

\begin{table*}
\centering
\begin{tabular}{l|l|l|l|l|l}
\noalign{\smallskip} \hline \hline \noalign{\smallskip}
\multicolumn{1}{c}{\textbf{GRB}} & \multicolumn{1}{|c}{$z$} &
\multicolumn{1}{|c}{\textbf{T$_{90}$} (sec)} &
\multicolumn{1}{|c}{$E_\gamma$ ($10^{52}$ erg)} &
\multicolumn{1}{|c}{$\mathcal{M}_c$ ($M_\odot$)} &
\multicolumn{1}{|c}{$h$ ($10^{-22}$)}   \\
 & & & & \ \ \ \ \ \tiny{$f=100\div1000$ Hz} &  \\
\hline
090426 &    2.61 &  \ \ \ 0.33 &        1.53$\pm$ 0.15 & (54.6$\pm$1.6)$\div$(5.4$\pm$0.2)      &  3.19$\div$0.32  \\
090205 &    4.7 &   \ \ \   1.54 &      1.88$\pm$ 0.19   & (91.2$\pm$2.7)$\div$(9.1$\pm$0.3)   &  5.79$\div0.58$  \\
080913 &    6.44 &  \ \ \   1.08 &      7.48$\pm$ 1.05   & (261$\pm$11)$\div$(26$\pm$1) & 29.5$\div$3.0\\
080516 &    3.2 &   \ \ \ 1.38 &        2.68$\pm$ 0.27 & (57$\pm$2)$\div$(5.6$\pm$0.2)     &  $3.11\div$0.31  \\
070724A &   0.46 &  \ \ \   0.27 &      0.46$\pm$ 0.05  & (6.6$\pm$0.2)$\div$(0.66$\pm$0.02)      &  0.266$\div$0.027  \\
070506  &   2.31 &  \ \ \   1.3 &       2.34$\pm$ 0.23 & (34$\pm$1)$\div$(3.4$\pm$0.1)      &  1.57$\div$0.16  \\
070429B &   0.9 &   \ \ \   0.25 &      0.95$\pm$ 0.10  &  (14.2$\pm$0.4)$\div$(1.42$\pm$0.04)      &  0.603$\div$0.060  \\
051221A &   0.55 &  \ \ \   0.9  &      2.04$\pm$ 0.19  & (8.1$\pm$0.2)$\div$(0.81$\pm$0.02)     &  0.328$\div$0.033  \\
050922C &   2.2 &   \ \ \   1.41 &      7.50$\pm$ 0.68  &  (44$\pm$1)$\div$(4.5$\pm$0.1)     &  2.48$\div$0.25  \\
050813  &   1.8 &   \ \ \   0.16 &      8.55$\pm$ 0.89  &  (68$\pm$2)$\div$(6.8$\pm$0.2)    &  5.56$\div$0.56  \\
050509B &   0.23 &  \ \ \   0.06 &      0.73$\pm$ 0.08  & (8.4$\pm$0.2)$\div$(0.84$\pm$0.03)     &  0.692$\div$0.070  \\
\noalign{\smallskip} \hline \noalign{\smallskip}
\end{tabular}
\caption{Results obtained  by coalescing binary systems and by the
SWIFT Catalogue: in column 1, 2, 3 and 4, the GRB data, derived
from the SWIFT Catalogue, are reported; in  column 5, the chirp
mass for each GRB in the range of frequency 100$\div$1000 Hz
(calculated by eq. \eqref{eq:mass}) is reported. In the last
column we have the amplitude of the GWs calculated by eq.
\eqref{eq:amplitude} }\label{tab:results}.
\end{table*}

\section{Results}
\label{quattro}

\begin{figure}[!h]
  \centering
  \includegraphics[scale=1]{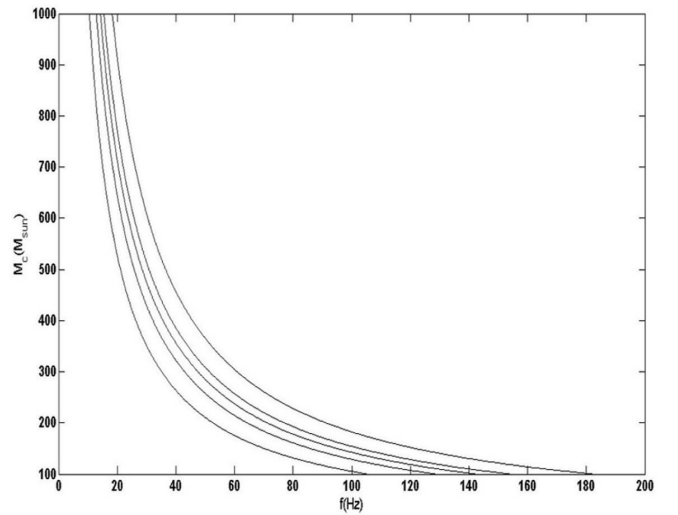}\\
  \caption{The chirp mass $\mathcal{M}_{c,\gamma}$ related to the various GRBs as function of the frequency $f$}
  \label{masschirpVSfreq}
\end{figure}

Using a sample of short GRBs of the SWIFT Catalogue for which the
redshift, the T$_{90}$ and the spectral index are measured (see
\cite{site}), we have calculate the peak energy with eq.
\eqref{correl} and the collimation-corrected energy with eq.
\eqref{logEgamma}. By eq.\eqref{eq:loss2}, we can obtain an
expression for the chirp mass $\mathcal{M}_c$ related to the rate
loss energy, the frequency and the redshift of the binary system:
\begin{equation}
 \mathcal{M}_c= \left( {\frac{{32}}{5}} \right)^{ - 3/20} \frac{{c^{3/2} }}{{G^{7/10} }}\frac{{\left( {1 + z} \right)^2 }}{{\pi f}}\left[ {\frac{{dE}}{{dt }}} \right]^{3/10}.
\end{equation}
Eq.\eqref{eq:mass} can be used to determine the chirp mass of the
binary coalescing systems, using redshifts of short GRBs. We
suppose the equivalence between T$_{90}$ and the coalescing time
($\tau$) for binary systems given by (\cite{Buonanno})
\begin{equation}
\tau  \simeq \left[ {\frac{{130}}{f}} \right]^{\frac{8}{3}} \left[ {\frac{{1.21M_ \odot  }}{M}} \right]^{\frac{5}{3}} {\rm{Hz}}^{\frac{8}{3}} {\rm{s}}.
\end{equation}
The results are reported in the following Table 3.
\begin{table}[!h]
\centering
\begin{tabular}{l|l}
\noalign{\smallskip} \hline \hline \noalign{\smallskip}
\multicolumn{1}{c}{$\mathcal{M}_c$ ($M_\odot$)} &
\multicolumn{1}{|c}{$\tau$ (sec)}   \\
 & \ \ \ \ \ \tiny{$f=100\div1000$ Hz}   \\
\hline
1.21 & (2$\div$0.043) \\
8.67 & (0.07$\div$1.6$\times10^{-4}$) \\
0.69 & (5.1$\div$0.01) \\
\noalign{\smallskip} \hline \noalign{\smallskip}
\end{tabular}
\caption{The coalescing time for NS-NS, BH-BH, WD-WD systems, respectively.}\label{tab:time}
\end{table}
Then we suppose  that the collimation-corrected energy is similar
to GW's energy emitted and  the frequency are tuned in the range
$100\div1000$ Hz (\cite{Formisano}), by  eq.\eqref{eq:mass} we can
write
\begin{equation}\label{eq:mass}
 \mathcal{M}_{c,\gamma}= \left( {\frac{{32}}{5}} \right)^{ - 3/20} \frac{{c^{3/2} }}{{G^{7/10} }}\frac{{\left( {1 + z} \right)^2 }}{{\pi f}}\left[ {\frac{{E_\gamma  }}{{T_{90} }}} \right]^{3/10}.
\end{equation}
The error on chirp mass is calculated by the error on the
collimation-corrected energy with the standard error propagation.
We suppose that the coalescence of binary systems is caused only
by the emission of GWs (neglecting, for example, mass exchanges
and neutrino emissions). Now, if we fix the cosmological model, we
can calculate, in principle,  the  luminosity distance for each
GRBs (\cite{Izzo}) by:
\begin{equation}\label{eq:DL}
D_l(z)=\int_0^z \frac{d \xi}{H(\xi)}\,.
\end{equation}
Considering the $\Lambda$CDM model and fixing as  priors
$H_0=70.5\pm1.3 $km/s/Mpc and $q_0=-0.57\pm0.17$ (\cite{Koma};
\cite{Virey}), we can calculate the distances and then we can use
them to estimate  the GW amplitude  by eq. \eqref{eq:amplitude}.
The error on the GW amplitudes can be assumed negligible. The
results are reported in Table \ref{tab:results}. In
Fig.\ref{masschirpVSfreq}, the qualitative behavior of the chirp
mass ${\mathcal{M}_{c,\gamma}}$ as a function of the frequency $f$
is reported for any GRB in the Table \ref{tab:results}. The shift
of the various curves is due to the ratio $E_{\gamma}/T_{90}$ and
to the redshift $z$.

\section{Conclusions}
\label{cinque}

In this paper, we have derived  chirp masses associated to a
sample of short GRBs (under the hypothesis that they are emitted
by binary systems whose redshifts can be estimated considering
them comparable  to the GRB redshift). In such a way, considering
the coalescing time equal to the T$_{90}$ and the energy lost
(during the final phase of the coalescing process) equal to the
emission of short GRBs, we have found that the chirp masses,
obtained by the simulation, are comparable to the theoretical
chirp masses associated to the coalescing binary systems. In
particular the range of the chirp masses is reliable  for BH-BH
coalescing systems, but for the GRB070724A, GRB070429B,
GRB051221A, GRB050509B the chirp masses is also reliable for NS-NS
and WD-WD coalescing systems. The masses of NSs are typically of
order $1.4M_\odot$. Stellar-mass BHs, as observed in X-ray
binaries, are in general more massive, typically with masses of
order $10M_\odot$, and therefore are expected to emit even more
powerful GW signals during their inspiraling and coalescing phases
(\cite{bulik}). The standard models  describing the GRB sources
are, in general,  gravitational collapse, or related to  fast
increases of the mass of one or more objects that involve an
enormous release of neutrinos and antineutrinos
(\cite{Meszaros:2006rc}; \cite{Schon}). As discussed, GRBs can be
rougly separated into two classes (\cite{Weeeks}), long GRBs (with
T$_{90}$ $\gtrsim2$s), associated to gravitational collapse of
very massive stars, and short GRBs (with T$_{90}$ $\lesssim2$s),
associated to mergers of compact objects. In particular we have
discussed  short GRBs, because a vast amount of energy that
coalescing binary systems (NS-NS, BH-BH or WD-WD) are thought to
release is comparable to the estimated energy release of GRBs
($\approx 10^{51}$ to $10^{53}$erg) thereby suggesting that
coalescing binary systems can be considered as possible source of
observed GRBs (\cite{Ramesh}). A binary system composed by two
objects is not stable because it loses energy by emitting GWs.
Such a loss of energy causes a progressive approach of the two
objects between them and so the coalescence of such objects. In
several cases, an accretion  disk  is produced at the rate of $0.1
M_\odot$s$^{-1} $ with an energy release of $\sim10^{53}$ erg
(\cite{eichler89}). On the other hand, the merging of such objects
can release up to $10 ^{53} $erg in $10$ms (\cite{eichler89}). A
meaningful fraction is issued in the form of neutrinos
$\sim10^{53}$ erg (\cite{eichler89}) and antineutrinos
($\sim10^{51}$ erg (\cite{eichler89})) that  annihilate by
producing electrons and positrons. The final result of such a
process is a sphere of energy that expands. Theoretical estimates
yield merger rates that can easily accommodate the observed burst
rate, with engine lifetimes and energy releases roughly consistent
with burst properties of cosmological populations. The locations
of short GRBs with respect to their host galaxies are compatible
with the kicks delivered by NS-NS binary systems at their birth
time. Short GRBs could be a primary source population for the
Laser Interferometer Gravitational Wave Observatory and other
ground-based GWs detectors. Next generation of interferometers (as
LISA (\cite{LISA}) or Advanced-VIRGO and LIGO) could play a
decisive role in order to detect GWs from these systems. At
advanced level, one expects to detect at least tens NS-NS
coalescing events per year, up to distances of order $2~$Gpc,
measuring the chirp mass with a precision better than $0.1\%$.

\end{document}